\documentclass[a4paper,twoside,twocolumn,english,prb,showpacs]{revtex4}
\usepackage[T1]{fontenc}
\usepackage[latin1]{inputenc}
\usepackage{graphicx}

\makeatletter

\usepackage{babel}
\makeatother
\begin{document}

\title{Resonance phenomena in asymmetric superconducting quantum interference
devices}

\begin{abstract}
Theory of self induced resonances in asymmetric two-junction interferometer
device is presented. In real devices it is impossible to have an ideal
interferometer free of imperfections. Thus, we extended previous theoretical
approaches introducing a model which contains several asymmetries:
Josephson current $\epsilon$, capacitances $\chi$ and dissipation
$\rho$ presented in an equivalent circuit. Moreover, non conventional
symmetry of the order parameter in high temperature superconducting
quantum interference devices forced us to include phase asymmetries.
Therefore, the model has been extended to the case of $\pi$-shift
interferometers, where a phase shift is present in one of the junctions. 
\end{abstract}

\author{T. P. Polak, E. Sarnelli}


\address{Consiglio Nazionale delle Ricerche - Istituto Nazionale per la Fisica
della Materia, Complesso Universitario Monte S. Angelo, 80126 Naples,
Italy}

\address{Istituto di Cibernetica ''E.Caianiello'' del CNR, Via Campi Flegrei
34, I-80078 Pozzuoli, Italy.}

\maketitle

\section{Introduction}

Superconducting quantum interference devices (SQUIDs) are the most
employed superconductive electronic circuits in practical applications.\cite{bergeala,fagaly,kouznetsov,yanga,wu,mosher,lindstrom,weinstock,gross}
With the discovery of high-temperature superconductors (HTS) also
high-temperature SQUIDs have been developed.\cite{barthel,kawasaki,koelle,wu1,lee}
This class of devices, although less sensitive than the most competitive
low-temperature SQUIDs, have been used in several applications, where
portability and/or positioning as much as high working temperatures
are needed. Moreover, the demonstration of an unconventional symmetry
of the order parameter in \emph{YBaCuO} (YBCO),\cite{sigrist,harlingen,tsuei}
opened new horizons for using the so-called pi-SQUIDs in superconductive
electronics. Indeed, $\pi$-SQUIDs\cite{schulz} can be used to self-frustrate
quantum bit circuits or to feed RSFQ (rapid single flux quantum) devices.\cite{ioffe,jan}%
\begin{figure}
\includegraphics[%
  scale=0.4]{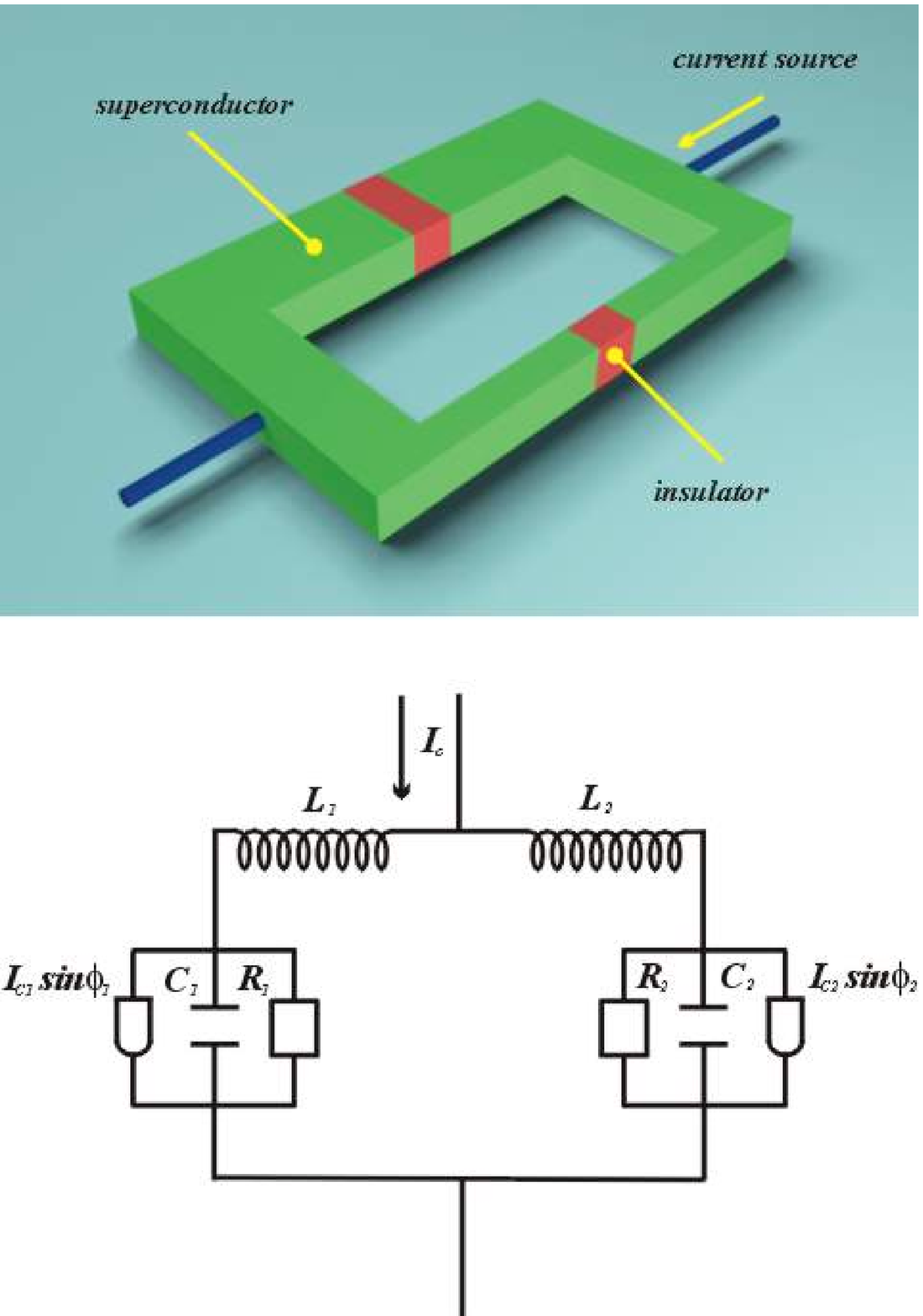}

\caption{Theoretical model of an asymmetric superconducting quantum interference
device and the equivalent circuit contains two Josephson junctions
with the critical current $I_{Ci}$ and parallel capacitance $C_{i}$.
Each single junction contains parallel linear resistance $R_{i}$
and the interferometer is fed by an external source $I_{c}$. The
self-inductances of the junctions are equal $L_{i}$ and $\phi_{i}$
is the phase difference across the $i$th junction.\label{theoretical-model}}
\end{figure}
 As a consequence, a full knowledge of properties of HTS SQUIDs is
at great importance. In particular, the aspects limiting their utilization
in applications have to be explored. We can consider two effects limiting
performance of HTS zero-or $\pi$-SQUIDs (zero indicates the conventional
SQUID where no phase shift has been established along the superconducting
loop): asymmetries in the junction properties and anomalous electrical
behavior induced by an arbitrary phase shift in one of the two junctions
forming the interferometer. Asymmetries in conventional low-temperature
devices have been first examined by Tesche and Clarke.\cite{tesche}
In their paper a complete study of the performance in terms of noise
characteristics has been carried out. The interest on asymmetric SQUIDs
grew up again after the discovery of HTS. Indeed, the parameter spread
in HTS SQUIDs is often so large that significant asymmetries arise.
Hence, it is particularly hard to fabricate two identical HTS Josephson
junctions, even though they are very close to each other on the chip.
Performance of asymmetric SQUIDs have been analyzed by Testa and co-workers.\cite{testa,testa1}
From their papers it is evident that higher magnetic sensitivities
are achieved when asymmetric SQUIDs are used. The asymmetry combined
with a damping resistance leads to a flux to voltage transfer coefficient
several times larger than the one typical for symmetric devices, together
with a lower magnetic flux noise. The large ratio of the flux to voltage
transfer coefficient allows a direct coupling to an external preamplifier
without the need of an impedance matching flux transformer or additional
positive feedback circuitry. This simplifies the read-out electronics,
as required in multi-channel systems for low-noise measurements. However,
the final performance of a dc-SQUID is influenced by the presence
of undesired anomalies occurring on the current-voltage (IV) characteristics,
namely Fiske or resonant steps.\cite{barone} Such structures originate
for the non-linear interaction between the resonant cavity, represented
by the superconducting loop, and an rf current component - the ac
Josephson current in the junctions. This system may be treated with
the equivalent electrical resonant circuit, as shown in Fig. 1.

A deep investigation of the properties of resonances in asymmetric
SQUIDs, also including different phase shift in the SQUID loop, is
mandatory and can be very useful for people involved in SQUID design.
Self-resonances occurring in superconducting interferometers are considered
to be phenomena reducing performance of high-sensitive SQUIDs. Indeed,
Zappe and Landman\cite{zappe} first investigated experimentally resonances
in low-Q Josephson interferometers. The analysis was taken again by
Tuckerman and Magerlein,\cite{tuckerman} who presented a theoretical
and experimental investigation of resonances in symmetric devices.
Successively, Faris and Walsamakis\cite{faris} showed characteristic
of resonances in asymmetric two-junction interferometers, introducing
an important distinction between current- and voltage-controlled cases.
Based on their analysis, Camerlingo et al.\cite{camerlingo} reported
an experimental work showing the effect of the loop capacitance on
resonant voltages in asymmetric interferometers. Recently, the nature
of resonances in SQUIDs in which a significant flux is coupled to
the Josephson junctions, called spatially distributed junctions (SDJ)
dc-SQUID, has been analyzed by Chesca.\cite{chesca} He showed that
useful information about the order parameter symmetry can be provided
by studying directly the magnetic field dependences of both the dc
Josephson critical current and self-induced resonant modes of dc-SQUIDs
made of non-conventional superconductors. The further analysis of
voltage states in current-voltage characteristics of symmetric dc
$\pi$-SQUIDs, in which the junctions are equal and not-distributed
circuital elements, has been done by Chesca and co-workers.\cite{chesca1}
Moreover, d-wave induced zero-field resonances in dc $\pi$-SQUIDs
have also been observed.\cite{chesca2} 

In our work we present a full investigation of resonances in asymmetric
SQUIDs, also in the presence of asymmetries in the junction phases.
The outline of the paper is the following: In Sec. II we outline the
model Hamiltonian, and we derive equations for asymmetric dc-SQUIDs.
In Sec III we present the method and assumptions which have been made.
Sec. IV we present our results considering special cases and their
relevance to the other theoretical works. Finally in Sec. V we discuss
the relevance obtained results to the experimental situations.

\section{Model}

We start with defining an asymmetric superconducting quantum interference
device (ASQUID) which consists of two Josephson junctions (see Fig.
\ref{theoretical-model}). Each of them has a critical current $I_{Ci}$
and a parallel capacitance $C_{i}$. We assume also that single junction
contains a parallel linear resistance $R_{i}$ and interferometer
is fed by an external source $I_{c}$, but the details of the equivalent
circuit will be specified later. The self-inductances of the junctions
in ASQUID are equal to $L_{1}$ and $L_{2}$. We do not consider mutual
inductances between the junctions. Hamiltonian of the ASQUID contains
three parts\cite{barone,likharev}:\begin{equation}
\mathcal{H=H}_{C}+\mathcal{H}_{J}+\mathcal{H}_{M}.\label{hamiltonian}\end{equation}
First term on the right side of Eq. (\ref{hamiltonian}) defines electrostatic
energy\begin{equation}
\mathcal{H}_{C}=\frac{1}{2}C_{1}V_{1}^{2}+\frac{1}{2}C_{2}V_{2}^{2},\end{equation}
where $V_{i}$ is the voltage across the $i$th junction. The last
equation can be transformed to the phase representation using the
Josephson relation $\dot{\phi}=2\pi/\Phi_{0}V$:\begin{equation}
\mathcal{H}_{C}=\frac{1}{2}\left(\frac{\Phi_{0}}{2\pi}\right)^{2}\left(C_{1}\dot{\phi}_{1}^{2}+C_{2}\dot{\phi}_{2}^{2}\right),\end{equation}
 where $\phi_{i}$ is a phase difference across the $i$th junction.
The second term is the Josephson energy: \begin{equation}
\mathcal{H}_{J}=E_{J,1}\left(1-\cos\phi_{1}\right)+E_{J,2}\left(1-\cos\phi_{2}\right),\end{equation}
where $E_{J,i}=\Phi_{0}/2\pi I_{C,i}$. To complete the set of equations
for the interferometer one should take into account that loop current
$I_{L}$ can contribute to the flux. The gauge invariant superconducting
phase differences between the edges of any loop and magnetic flux
are directly related by the fluxoid quantization relation:\begin{equation}
\phi_{2}-\phi_{1}=2\pi n+\phi_{ext}-\frac{2\pi}{\Phi_{0}}L_{+}I_{L},\end{equation}
where $n$ is an integer and\begin{eqnarray}
L_{+} & = & L_{1}+L_{2},\\
I_{L} & = & \frac{L_{1}I_{1}-L_{2}I_{2}}{L_{+}}.\end{eqnarray}
 Finally, for $n=0$ magnetic energy takes the form:\begin{equation}
\mathcal{H}_{M}=\frac{1}{2}L_{+}I_{L}^{2}=\frac{1}{2}\left(\frac{\Phi_{0}}{2\pi}\right)^{2}\frac{\left(\phi_{2}-\phi_{1}-\phi_{ext}\right)^{2}}{L_{+}}.\end{equation}
At this stage we do not provide an information about dissipative environment
and external forces which will be discussed later. Applying the Euler-Lagrange
equation \begin{equation}
dt\left(\partial_{\dot{\phi}_{n}}\mathcal{L}\right)-\partial_{\phi_{n}}\mathcal{L}=0\end{equation}
 to the Lagrangian\begin{eqnarray}
\mathcal{L} & = & \frac{1}{2}\left(\frac{\Phi_{0}}{2\pi}\right)^{2}\left(C_{1}\dot{\phi}_{1}^{2}+C_{2}\dot{\phi}_{2}^{2}\right)\nonumber \\
 &  & -\frac{1}{2}\left(\frac{\Phi_{0}}{2\pi}\right)^{2}\frac{\left(\phi_{2}-\phi_{1}-\phi_{ext}\right)^{2}}{L_{+}}\nonumber \\
 &  & -E_{J,1}\left(1-\cos\phi_{1}\right)-E_{J,2}\left(1-\cos\phi_{2}\right),\end{eqnarray}
we find equations of motion

\begin{eqnarray}
\left(\frac{\Phi_{0}}{2\pi}\right)^{2}C_{1}\ddot{\phi_{1}}+E_{J1}\sin\phi_{1} & = & \left(\frac{\Phi_{0}}{2\pi}\right)^{2}\frac{\left(\phi_{2}-\phi_{1}\right)}{L_{+}},\\
\left(\frac{\Phi_{0}}{2\pi}\right)^{2}C_{2}\ddot{\phi_{2}}+E_{J,2}\sin\phi_{2} & = & \left(\frac{\Phi_{0}}{2\pi}\right)^{2}\frac{\left(\phi_{1}-\phi_{2}\right)}{L_{+}}.\end{eqnarray}
 Similarly to Tesche\cite{tesche} we introduce the following parameters
\begin{eqnarray}
C_{1} & = & \left(1+\chi\right)C,\qquad C_{2}=\left(1-\chi\right)C,\\
E_{J,1} & = & \left(1+\epsilon\right)E_{J},\qquad E_{J,2}=\left(1-\epsilon\right)E_{J},\\
L_{1} & = & \left(1+\lambda\right)\frac{L}{2},\qquad L_{2}=\left(1-\lambda\right)\frac{L}{2},\end{eqnarray}
where dimensionless anisotropy quantities $\chi$, $\epsilon$ and
$\lambda$ describe the relative deviations of the model parameters
from the corresponding average values $C$, $E_{J}$ and $L$. We
can vary the values of the anisotropy parameters within the range
$\left[0,1\right)$, where zero leads to the isotropic model and value
one completely rules out presence of one junction from the interferometer.
Since $L_{+}=L$ we conclude that a difference between inductances
does not influence the dynamics of the model. After renormalization
to dimensionless quantities \begin{eqnarray}
\omega_{c}^{2} & = & \left(\frac{2\pi}{\Phi_{0}}\right)^{2}\frac{E_{J}}{C}=\frac{1}{LC},\\
\beta & = & \frac{2\pi}{\Phi_{0}}I_{C}L,\end{eqnarray}
we finally obtain two coupled non-linear second-order differential
equations describing an ASQUID:

\begin{eqnarray}
\left(1-\chi\right)\ddot{\phi_{1}}+\left(1-\epsilon\right)\sin\left(\phi_{1}+\vartheta_{1}\right) & = & \frac{\left(\phi_{2}-\phi_{1}\right)}{\beta},\label{first symmetry}\\
\left(1-\chi\right)\ddot{\phi_{2}}+\left(1+\epsilon\right)\sin\left(\phi_{2}+\vartheta_{2}\right) & = & \frac{\left(\phi_{1}-\phi_{2}\right)}{\beta}.\label{second symmetry}\end{eqnarray}

Until now we have not considered dissipation effects and specific
geometry of the circuit. First, we have to decide, what modes of operation
we think about: current controlled (CC) or voltage controlled (VC)?
This is a crucial point simply because a choice we make is going to
affect our system. For the VC case where SQUID is excited by a voltage
source $V_{s}$ we have to add terms proportional to $V_{s}t$ to
the equations. The difference caused by various excitation sources
affects frequencies of the oscillating modes of the system. In this
paper we assume that SQUID is current excited by a constant current
source (see Fig. \ref{theoretical-model}). This foundation leads
to an additional term $\gamma_{i}$ in both equations. Origin of the
last parameter is clear when we consider the equivalent loop of a
real interferometer\cite{tuckerman} where the center of the inductance
is fed by a gate current source $I_{g}$. Using notation from Tuckerman's
paper and the above information we can derive exact form of $\gamma_{i}$:\begin{equation}
\gamma_{1}=\frac{I_{g}+2I_{c}}{2I_{C}},\quad\gamma_{2}=\frac{I_{g}-2I_{c}}{2I_{C}}.\end{equation}
where $I_{c}$ is a circulating current. 

Considering dissipation due to a quasi-particle current we add parallel
resistances $R_{i}$. These dissipative currents flowing through the
junctions of the interferometer can vary from each other and, as a
consequence we have to introduce their asymmetry assuming\begin{equation}
\left(1+\rho\right)\alpha=\left(\frac{\Phi_{0}}{2\pi}\right)^{2}\frac{1}{R_{1}},\quad\left(1-\rho\right)\alpha=\left(\frac{\Phi_{0}}{2\pi}\right)^{2}\frac{1}{R_{2}}.\end{equation}

Different phase shift can be added to each junction separately putting
$\phi_{i}+\vartheta_{i}$ in Eq. (\ref{first symmetry}) and Eq. (\ref{second symmetry}).
We see that values $\vartheta_{0}=0$ and $\vartheta_{1}=\pi$ lead
to the opposite sign of the current which means its opposite direction.
Finally, we write the equations for ASQUID with phase shift in form:\begin{eqnarray}
\left(1+\chi\right)\ddot{\phi_{1}}+\left(1+\rho\right)\alpha\dot{\phi_{1}}+\left(1+\epsilon\right)\sin\left(\phi_{1}+\vartheta_{1}\right)\nonumber \\
=\gamma_{1}+\frac{\left(\phi_{2}-\phi_{1}\right)}{\beta}, &  & \textrm{}\label{diff1}\\
\left(1-\chi\right)\ddot{\phi_{2}}+\left(1-\rho\right)\alpha\dot{\phi_{2}}+\left(1-\epsilon\right)\sin\left(\phi_{2}+\vartheta_{2}\right)\nonumber \\
=\gamma_{2}+\frac{\left(\phi_{1}-\phi_{2}\right)}{\beta}.\label{diff2}\end{eqnarray}
 In order to obtain similar node equations one can also use Kirchoff's
current law to the specific circuit. We have to mention that the noise
effects are not present in our analysis. Choice of parameters $\chi=\epsilon=\rho=\vartheta=0$
stands for the fully symmetric case.

\section{Method}

We shall analyze two coupled differential equations (\ref{diff1})
and (\ref{diff2}) for the case $\beta\leq1$ that coupling between
the two junctions of the interferometer is strong and, hence the last
terms of the right hand in Eqs. (\ref{diff1}) and (\ref{diff2})
play important role since they contain expressions proportional to
$\pm\beta^{-1}\left(\phi_{2}-\phi_{1}\right)$. Let us introduce new
variables\begin{eqnarray}
\phi_{-} & = & \frac{\phi_{1}-\phi_{2}}{2},\qquad\phi_{+}=\frac{\phi_{1}+\phi_{2}}{2},\\
\gamma_{-} & = & \frac{\gamma_{1}-\gamma_{2}}{2},\qquad\gamma_{+}=\frac{\gamma_{1}+\gamma_{2}}{2},\\
\vartheta_{-} & = & \frac{\vartheta_{1}-\vartheta_{2}}{2},\qquad\vartheta_{+}=\frac{\vartheta_{1}+\vartheta_{2}}{2},\end{eqnarray}
where $\phi_{-}$ represents the flux number (this parameter distinguishes
interferometer from a point junction), and $\phi_{+}$ is the average
phase difference of the junctions. For the equivalent circuit of a
real interferometer $\gamma_{-}$ can be recognized as a control current
$I_{c}$ and $\gamma_{+}$ as a bias current $I_{g}$. Parameters
$\vartheta_{\pm}$ are relative changes of the phase shifts present
in each junction. In terms of the above we write equations (\ref{diff1})
and (\ref{diff2}) in form\begin{eqnarray}
\ddot{\phi_{+}}+\alpha\dot{\phi_{+}}+\sin\left(\phi_{+}+\vartheta_{+}\right)\cos\left(\phi_{-}+\vartheta_{-}\right)-\gamma_{+}\nonumber \\
+\chi\ddot{\phi_{-}}+\alpha\rho\dot{\phi_{-}}+\epsilon\sin\left(\phi_{-}+\vartheta_{-}\right)\cos\left(\phi_{+}+\vartheta_{+}\right)\nonumber \\
=0,\label{diff3}\end{eqnarray}
\begin{eqnarray}
\ddot{\phi_{-}}+\alpha\dot{\phi_{-}}+\sin\left(\phi_{-}+\vartheta_{-}\right)\cos\left(\phi_{+}+\vartheta_{+}\right)-\gamma_{-}+\frac{2}{\beta}\phi_{-}\nonumber \\
+\chi\ddot{\phi_{+}}+\alpha\rho\dot{\phi_{+}}+\epsilon\sin\left(\phi_{+}+\vartheta_{+}\right)\cos\left(\phi_{-}+\vartheta_{-}\right)\nonumber \\
=0.\label{diff4}\end{eqnarray}
In the following we have to assume a form of the solution. The voltage
variations appearing in ASQUID come from the interaction between the
junction current and circuit's elements. We assume voltage sinusoidal
variations with $\mathrm{dc}$ component $V$, $\mathrm{ac}$ amplitude
$v$, frequency $\omega$ and phase $\varphi$:\begin{equation}
V\left(t\right)=V+v\cos\left(\omega t+\varphi\right),\end{equation}
where other harmonics are filtered out. Using the Josephson relations
and integrating out we get for $i$th junction:\begin{equation}
\phi_{i}\left(t\right)=\phi_{0,i}+\omega t\pm\delta\sin\left(\omega t+\varphi\right),\end{equation}
where $\delta=\frac{v}{V}$. The flux number $\phi_{-}$ and the average
phase difference $\phi_{+}$ can be written:\begin{eqnarray}
\phi_{-} & = & \phi_{c}-\delta\sin\omega t,\\
\phi_{+} & = & n\omega t-\theta.\end{eqnarray}
 where $\phi_{c}$ is the average value of the internal phase $\phi_{-}$.
In order to account the difference between odd and even behavior of
the ASQUID interferometer we define:\begin{eqnarray}
\phi_{-} & = & \phi_{c}-\delta\sin\omega t-k\frac{\pi}{2},\label{zalozenie1}\\
\phi_{+} & = & n\omega t-\theta-k\frac{\pi}{2},\label{zalozenie2}\end{eqnarray}
where $k$ is equal $0\left(1\right)$ for even (odd) number of resonances.

\section{Results}

Substituting expressions (\ref{zalozenie1}) and (\ref{zalozenie2})
into equations (\ref{diff3}) and (\ref{diff4}) and extracting by
calculating average over time the $\mathrm{dc}$, $\sin\omega t$
and $\cos\omega t$ Fourier components we get:\begin{eqnarray}
\alpha n\omega & = & \gamma_{+}-J_{n}\left(\delta\right)\cos\left(\theta-\vartheta_{+}\right)\sin\left(\phi_{c}+\vartheta_{-}\right)\nonumber \\
 & + & \epsilon J_{n}\left(\delta\right)\sin\left(\theta-\vartheta_{+}\right)\cos\left(\phi_{c}+\vartheta_{-}\right),\label{jeden}\end{eqnarray}

\begin{eqnarray}
\gamma_{-} & = & \frac{2}{\beta}\phi_{c}-J_{n}\left(\delta\right)\sin\left(\theta-\vartheta_{+}\right)\cos\left(\phi_{c}+\vartheta_{-}\right)\nonumber \\
 & + & \alpha\rho n\omega+\epsilon J_{n}\left(\delta\right)\cos\left(\theta-\vartheta_{+}\right)\sin\left(\phi_{c}+\vartheta_{-}\right),\label{dwa}\end{eqnarray}
\begin{eqnarray}
-\chi\delta\omega^{2} & = & J_{n}^{-}\left(\delta\right)\cos\left(\theta-\vartheta_{+}\right)\cos\left(\phi_{c}+\vartheta_{-}\right)\nonumber \\
 & + & \epsilon J_{n}^{-}\left(\delta\right)\sin\left(\theta-\vartheta_{+}\right)\sin\left(\phi_{c}+\vartheta_{-}\right),\label{trzy}\end{eqnarray}
\begin{eqnarray}
\alpha\rho\delta\omega & = & -J_{n}^{+}\left(\delta\right)\sin\left(\theta-\vartheta_{+}\right)\cos\left(\phi_{c}+\vartheta_{-}\right)\nonumber \\
 & + & \epsilon J_{n}^{+}\left(\delta\right)\cos\left(\theta-\vartheta_{+}\right)\sin\left(\phi_{c}+\vartheta_{-}\right),\label{cztery}\end{eqnarray}
\begin{eqnarray}
\delta\left(\frac{2}{\beta}-\omega^{2}\right) & = & J_{n}^{-}\left(\delta\right)\sin\left(\theta-\vartheta_{+}\right)\sin\left(\phi_{c}+\vartheta_{-}\right)\nonumber \\
 & + & \epsilon J_{n}^{-}\left(\delta\right)\cos\left(\theta-\vartheta_{+}\right)\cos\left(\phi_{c}+\vartheta_{-}\right),\label{piec}\end{eqnarray}
\begin{eqnarray}
\alpha\delta\omega & = & J_{n}^{+}\left(\delta\right)\cos\left(\theta-\vartheta_{+}\right)\sin\left(\phi_{c}+\vartheta_{-}\right)\nonumber \\
 & - & \epsilon J_{n}^{+}\left(\delta\right)\sin\left(\theta-\vartheta_{+}\right)\cos\left(\phi_{c}+\vartheta_{-}\right),\label{szesc}\end{eqnarray}
where \begin{equation}
J_{n}^{\pm}\left(\delta\right)=J_{n-1}\left(\delta\right)\pm J_{n+1}\left(\delta\right)\end{equation}
and $J_{n}\left(\delta\right)$ is the Bessel function of the first
kind.\cite{abramovitz} Using Eq. (\ref{jeden}), Eq. (\ref{szesc})
and Bessel function identity, we obtain\begin{eqnarray}
\alpha n\omega & = & \gamma_{+}-\frac{J_{n}\left(\delta\right)}{J_{n}^{+}\left(\delta\right)}\alpha\delta\omega\nonumber \\
 & = & \gamma_{+}-\frac{\alpha\delta^{2}\omega}{2n}.\end{eqnarray}
We define normalized excess current due to the resonance\begin{equation}
I_{exc}=\frac{\alpha\delta^{2}\omega}{2n}.\label{excess}\end{equation}
Above equations can be rewritten using the dimensionless damping parameter
$\Gamma\equiv\left(\alpha\omega_{r}\right)^{-1}$, where $\omega_{r}$
is the resonant frequency. Gamma was introduced by Werthamer\cite{werthamer}
and described the strength of the coupling of the current to the resonance
in case of the junction coupled to cavity. Several authors used it
as a damping parameter.\cite{tuckerman,zappe} We can combine equations
(\ref{dwa}) and (\ref{cztery}) :\begin{equation}
\gamma_{-}=\frac{2}{\beta}\phi_{c}+\alpha\rho n\omega+\frac{\alpha\rho\delta^{2}\omega}{2n},\end{equation}
and using relation for excess current we get\begin{equation}
\gamma_{-}=\frac{2}{\beta}\phi_{c}+\rho\gamma_{+}.\end{equation}
We see that formula (\ref{excess}) derived for the excess current
is universal in such sense that it holds even for asymmetric SQUID.
This expression is also true in the presence of any changes of the
phase shift in one of the junctions of the interferometer. We can
add squares of expressions (\ref{trzy}-\ref{szesc}): \begin{eqnarray}
\left[\frac{\delta\left(1-\tilde{\omega}^{2}\right)}{J_{n}^{-}\left(\delta\right)}\right]^{2}+\left[\frac{\alpha\delta\tilde{\omega}}{J_{n}^{+}\left(\delta\right)}\right]^{2}\nonumber \\
+\left[\frac{\chi\delta\tilde{\omega}^{2}}{J_{n}^{-}\left(\delta\right)}\right]^{2}+\left[\frac{\alpha\rho\delta\tilde{\omega}}{J_{n}^{+}\left(\delta\right)}\right]^{2} & = & 1+\epsilon^{2},\label{ancurrentG}\end{eqnarray}
where $\tilde{\omega}=\omega/\omega_{r}=V/V_{r}$ is the normalized
voltage. From above and Eq. (\ref{excess}) we can derive normalized
excess current dependence on voltage for given anisotropic parameters.
However analysis is complex and is better to simplify our model considering
special cases which could give us more insight into structure of resonances
in ASQUID.

\subsection{Special cases }

For a general choice of parameters equations (\ref{jeden})-(\ref{szesc})
are coupled and must be solved numerically. However considerations
of special cases can provide more insights into general solution of
the problem.

\subsubsection{Asymmetry of the Josephson current $\left(\epsilon\neq0\right)$}

\begin{figure}
\includegraphics[%
  scale=0.65]{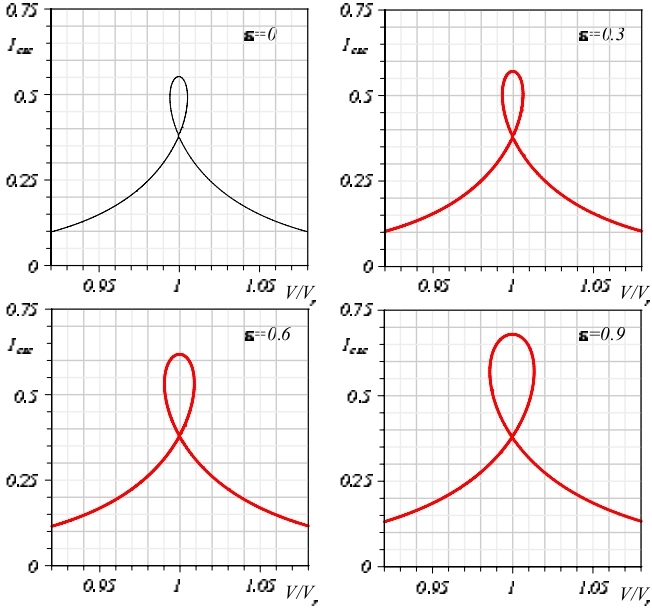}

\caption{Current voltage characteristics with Josephson current anisotropy
$\epsilon$, first resonance, $\Gamma=20$, $\beta=0.1$. Black color
of the curves used in this and next plots indicates the symmetric
SQUID $\chi=\epsilon=\rho=0$.\label{anIVeall}}
\end{figure}
\begin{figure}
\includegraphics[%
  scale=0.65]{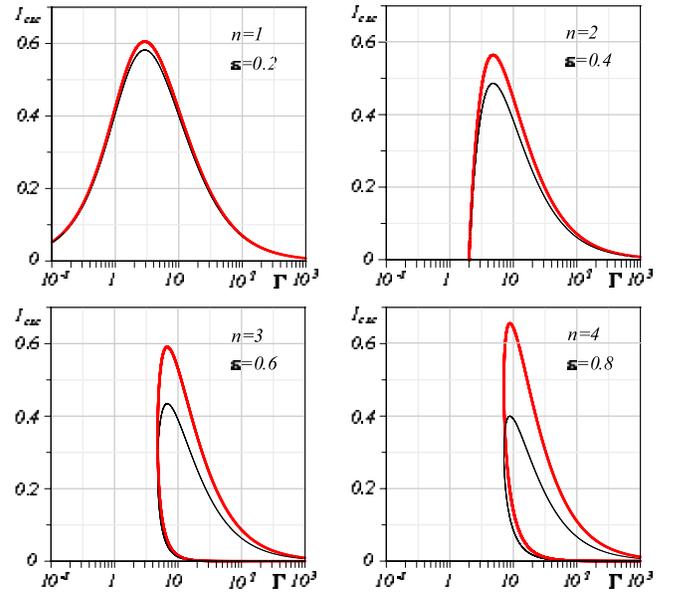}

\caption{The normalized resonant current $I_{exc}$ versus damping parameter
$\Gamma$ for different values of the Josephson current anisotropy
parameters $\epsilon$ (red curves), $n$th resonance.\label{anig1reall}}
\end{figure}
In that case we assume that only Josephson current asymmetry is present.
Then the Eq. (\ref{ancurrentG}) can be reduced to form\begin{equation}
\left[\frac{\delta\left(1-\tilde{\omega}^{2}\right)}{J_{n}^{-}\left(\delta\right)}\right]^{2}+\left[\frac{\alpha\delta\tilde{\omega}}{J_{n}^{+}\left(\delta\right)}\right]^{2}=1+\epsilon^{2}.\end{equation}
From the above expression coupled with Eq. (\ref{excess}) we can
derive the normalized current dependence on normalized voltage plots
with $\epsilon$ asymmetry (see Fig. \ref{anIVeall}). Also the normalized
resonant current versus damping parameter for several resonances can
be obtained (see Fig. \ref{anig1reall}).

\subsubsection{Asymmetry of the capacitances $\left(\chi\neq0\right)$ and resistances
$\left(\rho\neq0\right)$ with phase shift $\left(\vartheta_{\pm}\neq0\right)$}

Let us consider case when $\epsilon=0$ which means that asymmetry
of the Josephson current is not present. In this case equations (\ref{jeden})-(\ref{szesc})
are reduced to\begin{eqnarray}
\alpha n\omega & = & \gamma_{+}-J_{n}\left(\delta\right)\cos\left(\theta-\vartheta_{+}\right)\sin\left(\phi_{c}+\vartheta_{-}\right),\end{eqnarray}

\begin{eqnarray}
\gamma_{-} & = & \frac{2}{\beta}\phi_{c}-J_{n}\left(\delta\right)\sin\left(\theta-\vartheta_{+}\right)\cos\left(\phi_{c}+\vartheta_{-}\right)\nonumber \\
 &  & +\alpha\rho n\omega,\end{eqnarray}
\begin{eqnarray}
-\chi\delta\omega^{2} & = & J_{n}^{-}\left(\delta\right)\cos\left(\theta-\vartheta_{+}\right)\cos\left(\phi_{c}+\vartheta_{-}\right),\label{rown rezonansowe1}\end{eqnarray}
\begin{eqnarray}
\alpha\rho\delta\omega & = & -J_{n}^{+}\left(\delta\right)\sin\left(\theta-\vartheta_{+}\right)\cos\left(\phi_{c}+\vartheta_{-}\right),\label{rown rezonansowe2}\end{eqnarray}
\begin{eqnarray}
\delta\left(\frac{2}{\beta}-\omega^{2}\right) & = & J_{n}^{-}\left(\delta\right)\sin\left(\theta-\vartheta_{+}\right)\sin\left(\phi_{c}+\vartheta_{-}\right),\label{rown rezonansowe3}\end{eqnarray}
\begin{eqnarray}
\alpha\delta\omega & = & J_{n}^{+}\left(\delta\right)\cos\left(\theta-\vartheta_{+}\right)\sin\left(\phi_{c}+\vartheta_{-}\right).\label{rown rezonansowe4}\end{eqnarray}
First we will analyze low-$\Gamma$ case in order to compare our results
with the original calculations presented in literature.\cite{zappe,gueret}

\subsubsection*{Low-$\Gamma$ devices with symmetric values of the Josephson current
$\left(\epsilon=0\right)$}

\begin{figure}
\includegraphics[%
  scale=0.7]{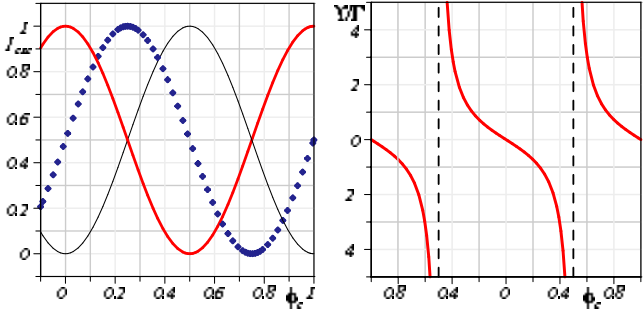}

\caption{Dependence of the normalized resonant current $I_{exc}$ with relative
phase shift $\vartheta_{-}=0.5$ (red), $0.25$ (dot), $0$ (black)
and the ratio $\Upsilon/\Gamma$ with no phase shift $\vartheta_{-}=0$
versus magnetic field $\phi_{c}$.\label{current phase} }
\end{figure}

At the resonance frequency $\omega=n\omega_{r}$ Eq. (\ref{rown rezonansowe3})
is satisfied when $\theta=\vartheta_{+}$. This condition rules out
equations with terms proportional to $\sin\left(\theta-\vartheta_{+}\right)$
and therefore there is no trace of the asymmetries of the Josephson
current $\epsilon$ and dissipation $\rho$. For small gamma $\Gamma$
devices $J_{n}^{\pm}\left(\delta\right)=0$ for $n>1$ and, hence
only the first resonance exists. We can derive the following equations

\begin{eqnarray}
-\frac{\delta}{\Upsilon} & = & \cos\left(\phi_{c}+\vartheta_{-}\right),\end{eqnarray}
\begin{eqnarray}
\frac{\delta}{\Gamma} & = & \sin\left(\phi_{c}+\vartheta_{-}\right).\end{eqnarray}
 where $\Upsilon\equiv\left(\chi\omega^{2}\right)^{-1}$ is dimensionless
parameter. Rearranging the last equation and putting into expression
for excess current we get:\begin{equation}
I_{exc}=\Gamma\sin^{2}\left(\phi_{c}+\vartheta_{-}\right),\end{equation}
which is general result for different SQUIDs. We can calculate other
relations:\begin{eqnarray}
I_{exc} & = & \frac{\Upsilon^{2}}{\Gamma}\cos^{2}\left(\phi_{c}+\vartheta_{-}\right),\\
\frac{\Upsilon}{\Gamma} & = & -\tan\left(\phi_{c}+\vartheta_{-}\right).\end{eqnarray}
which are plotted in Fig. \ref{current phase}. We see that the results
obtained previously by other authors\cite{zappe,gueret} are presented
in framework of our rather general calculations and can be derived
as special cases.

\subsubsection*{Analysis for not small $\Gamma$}

\begin{figure}
\includegraphics[%
  scale=0.65]{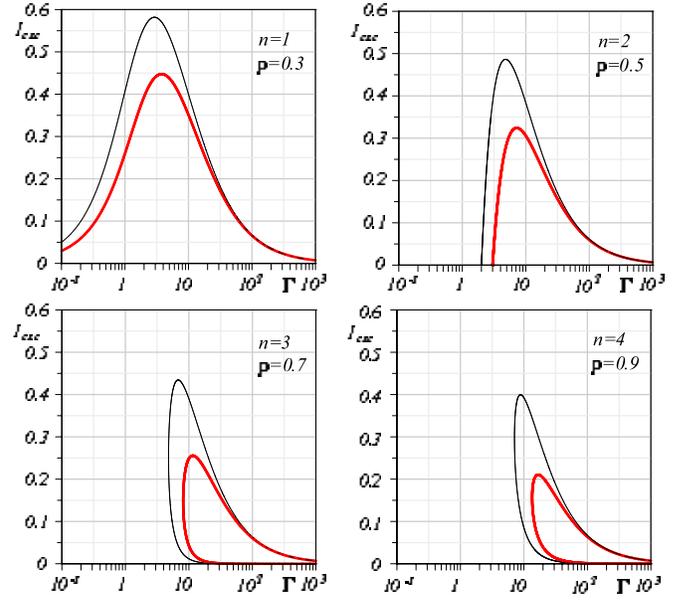}

\caption{The normalized resonant current $I_{\mathrm{exc}}$ versus damping
parameter $\Gamma$ for different values of the dissipation anisotropy
parameters $\rho$ (red curves), $n$th resonance.\label{anigrall}}
\end{figure}
\begin{figure}
\includegraphics[%
  scale=0.65]{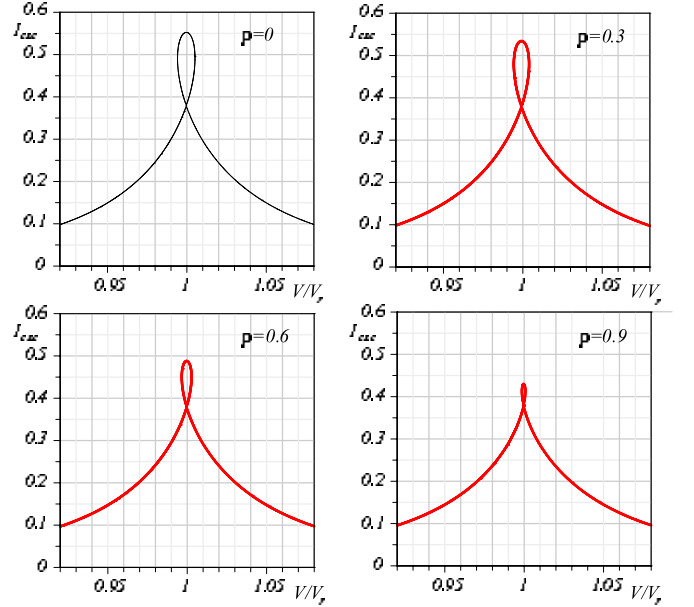}

\caption{Current voltage $\left(I_{\mathrm{exc}}-V/V_{r}\right)$ characteristics
with dissipation anisotropy $\rho$, first resonance $\left(n=1\right)$,
$\Gamma=20$, $\beta=0.1$.\label{anIVroall}}
\end{figure}

When $\Gamma$ is not small we cannot simplify equations using condition
under which Bessel functions can be approximated by zero except the
case of the first resonance. Putting $\epsilon=0$ in Eq. (\ref{ancurrentG})
we obtain:\begin{eqnarray}
\left[\frac{\delta\left(1-\tilde{\omega}^{2}\right)}{J_{n}^{-}\left(\delta\right)}\right]^{2}+\left[\frac{\alpha\delta\tilde{\omega}}{J_{n}^{+}\left(\delta\right)}\right]^{2}\nonumber \\
+\left[\frac{\chi\delta\tilde{\omega}^{2}}{J_{n}^{-}\left(\delta\right)}\right]^{2}+\left[\frac{\alpha\rho\delta\tilde{\omega}}{J_{n}^{+}\left(\delta\right)}\right]^{2} & = & 1.\label{ancurrent}\end{eqnarray}
From the above equation and expression (\ref{excess}) for the excess
current we can derive the normalized current voltage characteristics
for ASQUID. The second and the fourth terms of above equation can
be combined. We see that influence of the anisotropy of the dissipative
current\begin{equation}
\left[\frac{\alpha\delta\tilde{\omega}}{J_{n}^{+}\left(\delta\right)}\right]^{2}\rightarrow\left[1+\rho\right]\left[\frac{\alpha\delta\tilde{\omega}}{J_{n}^{+}\left(\delta\right)}\right]^{2}\end{equation}
\begin{figure}
\includegraphics[%
  scale=0.65]{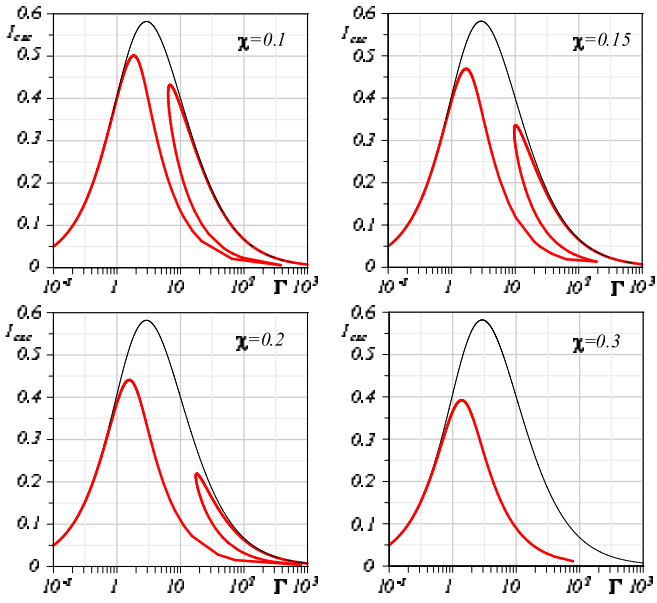}

\caption{The normalized resonant current $I_{\mathrm{exc}}$ versus damping
parameter $\Gamma$, different capacitance anisotropy parameter $\chi$,
first resonance $\left(n=1\right)$. \label{anig1rcall}}
\end{figure}
\begin{figure}
\includegraphics[%
  scale=0.65]{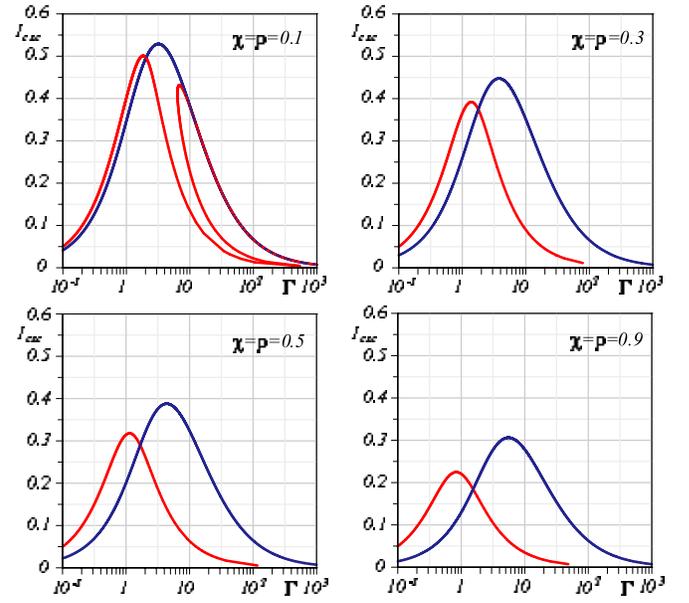}

\caption{The normalized resonant current $I_{\mathrm{exc}}$ versus damping
parameter $\Gamma$, for different values of the capacitance and dissipation
anisotropy parameters $\chi\left(red\right)=\rho\left(blue\right)$,
first resonance $\left(n=1\right)$.\label{anig1rrcall}}
\end{figure}
 manifests by the decreasing of the maximum value of the resonant
current, for given $n$th resonance mode, when we increase the anisotropy
parameter $\rho$ (see Fig. \ref{anigrall} and Fig. \ref{anIVroall}).
We observe a shift of the maximum value of $I_{exc}$ toward higher
values of the damping parameter $\Gamma$. Analysis of the influence
of the anisotropy of the capacitances can be done in the same manner.
We can again merge first and third terms of the Eq. (\ref{ancurrent}).
Contrary to previous simple case present one is more complex merely
because we have taken into account element proportional to $\tilde{\omega}^{4}$
which produces minor changes (see Fig. \ref{anig1rcall} and Fig.
\ref{anig1rrcall}). Now even small deviations of the anisotropy parameter
$\chi$ from equilibrium have a major impact on equations and in consequence
on behavior of the ASQUID. For small values of $\chi$, at fixed value
of the damping parameter $\Gamma$ there are two possible solutions
even for the first resonance. In symmetric SQUIDs this situation was
present for higher resonances $n\geq3$. Explanation of the latter
comes from the fact that the resonant circuit oscillates at a frequency
of $\omega_{r}$, while Josephson current in the junctions oscillates
at $n\omega_{r}$. In ASQUID we have three natural frequencies\cite{faris}
$\omega_{1,2}=\left(L_{+}C_{1,2}\right)^{-1/2}$ and related $\omega_{3}^{2}=\omega_{1}^{2}+\omega_{2}^{2}$
which can be excited by the $\mathrm{ac}$ Josephson effect and converted
through nonlinear interactions between junction and resonant circuit
into $\mathrm{dc}$ current steps. Therefore introducing capacitance
anisotropy we are able to create higher modes multivalued behavior
of the excess current even for the first resonance.

\begin{figure}
\includegraphics[%
  scale=0.65]{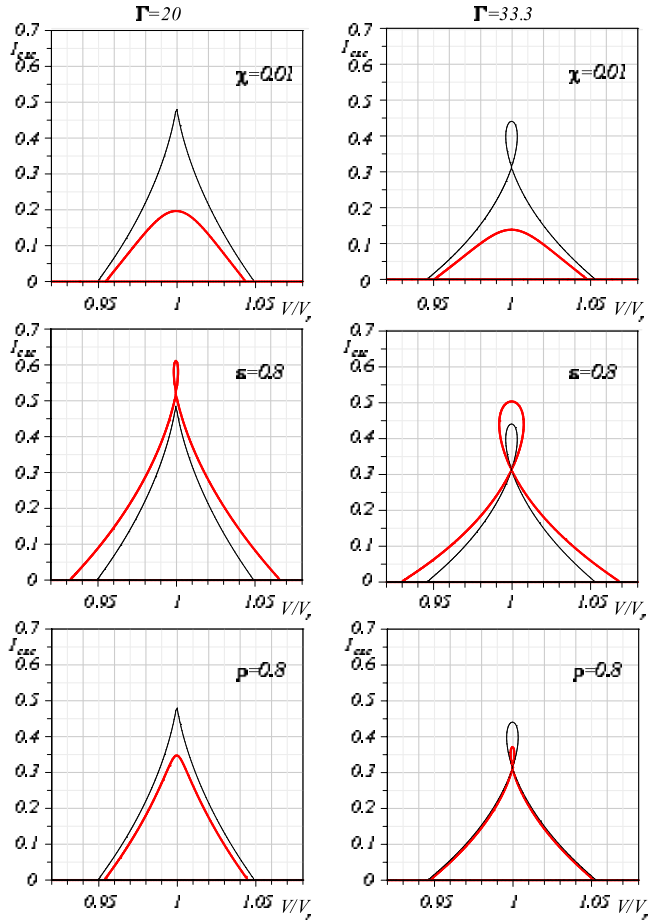}

\caption{The normalized resonant current $I_{\mathrm{exc}}$ versus normalized
voltage $V/V_{r}$, with different values of the anisotropy parameters,
capacitance $\chi$, the Josephson current $\varepsilon$, and dissipation
$\rho$ for second resonance ($n=2$). Black curves refer to absence
of anisotropy parameters.}
\end{figure}

\subsubsection{Symmetric case $\left(\chi=\epsilon=\rho=0\right)$ with phase shift
$\left(\vartheta_{\pm}\neq0\right)$}

\begin{figure}
\includegraphics[%
  scale=0.4]{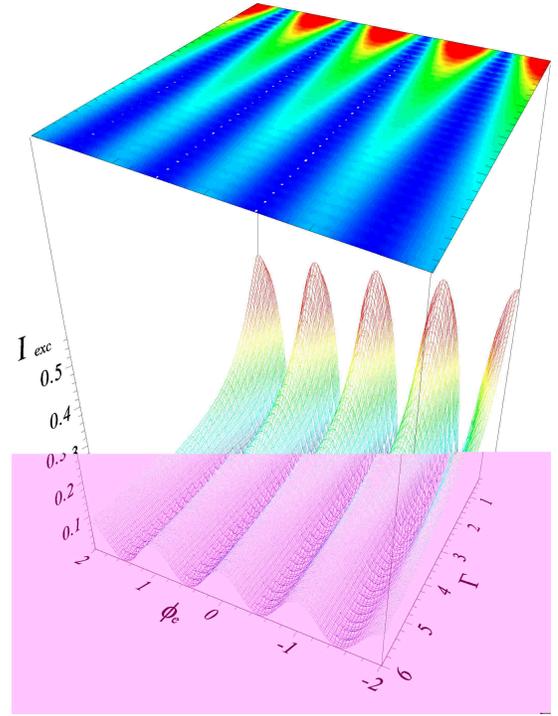}

\caption{Excess current $I_{exc}$ versus magnetic field $\phi_{e}$ (first
resonance, $n=1$) characteristic for different values of the damping
parameter $\Gamma$ for $0-\pi$ interferometer.\label{fig3d}}
\end{figure}

This case corresponds with a situation where different phase shift
is present in the junctions of the interferometer and analysis is
similar to one carried by Chesca.\cite{chesca} The equations take
form:

\begin{eqnarray}
\alpha n\omega & = & \gamma_{+}-J_{n}\left(\delta\right)\cos\left(\theta-\vartheta_{+}\right)\sin\left(\phi_{c}+\vartheta_{-}\right),\end{eqnarray}
\begin{eqnarray}
\gamma_{-} & = & \frac{2}{\beta}\phi_{c}-J_{n}\left(\delta\right)\sin\left(\theta-\vartheta_{+}\right)\cos\left(\phi_{c}+\vartheta_{-}\right),\end{eqnarray}
\begin{eqnarray}
\delta\left(\frac{2}{\beta}-\omega^{2}\right) & = & J_{n}^{-}\left(\delta\right)\sin\left(\theta-\vartheta_{+}\right)\sin\left(\phi_{c}+\vartheta_{-}\right),\label{piec shift}\end{eqnarray}
\begin{eqnarray}
\alpha\delta\omega & = & J_{n}^{+}\left(\delta\right)\cos\left(\theta-\vartheta_{+}\right)\sin\left(\phi_{c}+\vartheta_{-}\right).\label{szesc shift}\end{eqnarray}
We do not expect any changes in excess current - voltage characteristics.
Rather, as it was pointed out by Chesca the difference between SQUIDs
with various phase shifts can be visible only in magnetic field. In
order to calculate excess current dependence on magnetic field we
add squares of the equations (\ref{piec shift}) and (\ref{szesc shift}).
The resonant current is maximized when $\theta=\vartheta_{+}$ and
we can write the solution in parametric form:\begin{equation}
\left[I_{exc};\sin\left(\pi\phi_{e}+\vartheta_{-}\right)\right]=\left[\frac{\delta^{2}}{2\Gamma n};\frac{\delta}{\Gamma J_{n}^{+}\left(\delta\right)}\right].\end{equation}
where $\delta$ is a dummy variable. Changing value of the parameter
$\vartheta_{-}$ from $0$ to $-\pi/2$ we have $0-0$ and $0-\pi$
SQUID respectively. The shape of the surface describe current magnetic
field dependence (see Fig. \ref{fig3d}) remains unchanged but is
translated by a vector $\left[0;-\vartheta_{-}\right]$ along $\phi_{e}$
axis.

\section{Discussion}

\begin{figure}
\includegraphics[%
  scale=0.65]{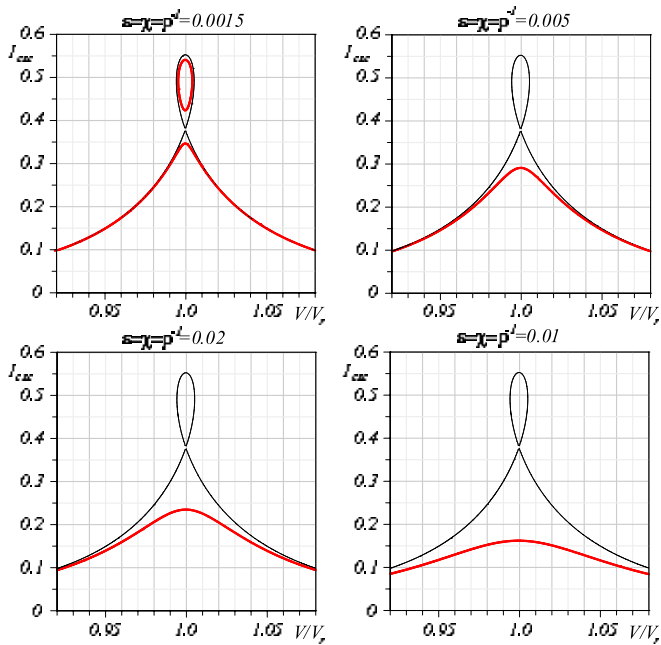}

\caption{Current voltage characteristics $\left(I_{\mathrm{exc}}-V/V_{r}\right)$
for several asymmetric configurations of the SQUIDs related to the
Ambegaokar-Baratoff formula ($\Delta I_{C}\sim\Delta C\sim\Delta R^{-1}$)
for changes of the junction area $\Delta A$, first resonance $n=1$,
$\Gamma=20$, $\beta=0.1$.\label{anIVerochiall}}
\end{figure}

The resonances in SQUIDs are investigated theoretically with several
asymmetries: Josephson current $\epsilon$, dissipation $\rho$ and
capacitance $\chi$. In real devices it is impossible to have an ideal
interferometer free of imperfections. In practice various deviations
of the interferometer parameters from average values can occur together
and mutually conceal each other. At this stage we have to separate
discussion related to low- and high-$T_{C}$ SQUIDs. In the former
case, experimentally, we are able to control asymmetry of dissipative
parameter $\rho$ adding a parallel resistor to the junction but it
is difficult to change the Josephson current independently from the
capacitance. To produce the asymmetry of the Josephson current in
the interferometer we can change the area of the junction $A$ or
thickness of the barrier $d$. Parallel-plate capacitor with area
$A$ of the plates and space $d$ between them has the capacitance
equal $C=\epsilon_{r}\epsilon_{0}A/d$ for $A\gg d^{2}$, where $\epsilon_{r}$
is the relative dielectric constant of the interlayer dielectric and
$\epsilon_{0}$ is the vacuum electric constant. On the other hand
the critical current can be written as $I_{C}=j_{C}A$ where $j_{C}$
is the critical current density. These two simple relations imply
that varying area $\Delta A$ of the junction in the interferometer
we change both capacitance and critical current proportionally $\Delta I_{C}\sim\Delta C$
at the same time. When no further resistor is added to the junctions
not only capacitance and Josephson current are related. From Ambegeokar-Baratoff\cite{ambegaokar}
formula we know that the product $I_{C}R_{N}$, where $R_{N}$ is
the resistance in normal state, has an invariant value which depends
only on the material in fixed temperature. Thus changing the value
of the Josephson current we alter the resistance of the junction.
Recapitulating these rather simple considerations we can introduce
asymmetry in the Josephson current changing the area of the junction
($\Delta I_{C}\sim\Delta C\sim\Delta R^{-1}$). Setting parallel resistor
we can control value of the resistance and vary dissipative parameter
independently from the current asymmetry. We can also imagine junctions
with different thicknesses of the barrier but technologically this
case is difficult to achieve thus we do not consider it. In experiments
with ASQUID both technically reached asymmetric cases do not differ
very much because of the capacitance anisotropy. As we see from Fig.
\ref{anIVerochiall} the biggest impact on the maximum value of the
resonant current has the anisotropy of the capacitance. Even small
changes of $\chi$ can decrease excess current almost to zero.

The situation changes completely when high-$T_{C}$ SQUIDs are considered.
On one hand, the probability to find junction parameter asymmetries
is particularly high, because high-$T_{C}$ junctions are intrinsically
affected by defects, as for instance faceting and/or oxygen vacancies
inside the barrier. Moreover, up to now, the charge transport process
is not completely understood, although various hypothesis have been
proposed,\cite{gross1,sarnelli,mennema} and other recent experiments
are still in progress.\cite{schneider,schulz} In particular, the
simple rule $I_{C}R_{N}=\mathrm{const}$ valid for low-$T_{C}$ SQUIDs
does not apply in the case of high-$T_{C}$ interferometers typically
used in applications, based on the symmetric bicrystal c-axis $\left[001\right]$
devices, and changing one single parameter is now possible. In such
interferometers, $I_{C}R_{N}$ is proportional to the critical current
density $J_{C}$ at low values and stays roughly constant at high-$J_{C}$
values.\cite{sarnelli1,hilgenkamp1} Moreover, HTS junctions are intrinsically
shunted and SQUIDs are fabricated with no additional shunt resistor.
As a consequence, the way to fabricate HTS SQUIDs with symmetric junctions
is probably to reduce junctions' widths, limiting the effect of the
interface defects. In all other cases, asymmetries will be very probable
and our analysis could be relevant to understand the presence of resonance
steps.

Different approach is necessary in the case of asymmetric $\left[001\right]$
or $\left[100\right]$ HTS bicrystal junctions, where the relation
$I_{C}R_{N}$ seems to be similar to the one of low-$T_{C}$ systems\cite{sarnelli1}
and the necessity to account for effects of a non-conventional symmetry
of the order parameter forces to include also the phase asymmetries
in studying dynamical states in HTS interferometers. Finally, also
the inclusion of the second harmonic term in the Josephson current
in order to account for experimental results\cite{ilichev,gardiner,lindstrom1}
is mandatory. This will be the argument of a separate paper, and the
possibility to deal with one single asymmetric parameter is now eventual.
Moreover, a non conventional symmetry of the order parameter forces
to include also phase asymmetries in studying dynamic states in high-$T_{C}$
interferometers. In this frame the calculations derived in the present
paper allow to investigate SQUID dynamics in both low- and high-$T_{C}$
asymmetric devices.

\section{Summary}

In this paper we have presented a detailed theoretical study of the
resonances in the asymmetric superconducting quantum interference
device. Analytical approach revealed the nature of the resonances
in the presence of several asymmetries: Josephson current $\epsilon$,
capacitances $\chi$ and dissipation $\rho$. Also we were able to
derive magnetic field dependence of the excess current in presence
of the magnetic field and phase shift. Our calculations imply that
deviations of the capacitances from the average value in SQUID have
profound impact on physics of the system. We have found that our theory
can be useful to determine asymmetry parameters present in lightly
damped ASQUIDs. Especially for SQUIDs produced from HTS materials
where deviations from average values are practically inevitable our
considerations are very helpful.

\begin{acknowledgments}
Authors would like to thank Prof. Antonio Barone and Dr. Ciro Nappi
for a lot of fruitful discussions. This work was supported by the
TRN {}``DeQUACS''.
\end{acknowledgments}

\end{document}